\documentclass{ws-procs975x65}

\begin{document}

\title{MONDIAN DARK MATTER, ENTROPIC GRAVITY, AND 
INFINITE STATISTICS}

\author{Y. JACK NG}

\address{Institute of Field Physics, Department of Physics \& 
Astronomy\\
Chapel Hill, NC 27599-3255, USA\\
E-mail: yjng@physics.unc.edu}

\begin{abstract}
We propose the concept of MoNDian dark matter 
which behaves like cold dark matter at
cluster and cosmic scales but emulates modified
Newtonian dynamics at the galactic scale. 
The connection between global physics
and local galactic dynamics is implemented 
via entropic gravity. We also 
give an alternative formulation of
MoNDian dark matter by using
an effective gravitational Born-Infeld theory.
In the latter approach, we show that the quanta
of MoNDian dark matter obey infinite
statistics. 
\end{abstract}


\bodymatter

\section{Introduction and Summary}
This talk is based on work\cite{HMN} done in collaboration
with C.M. Ho and D. Minic.\footnote{Supported in part by the US 
Department of Energy.}
As the first part of the title indicates,
the subject matter has to do with dark matter and modified Newtonian 
dynamics 
(MoND).  Rarely does one hear these two disparate things mentioned in
the same sentence.  But Ho, Minic and I think that it is natural to
put them together: in a real sense, they are dual to each other.
\footnote{We note that M. Kaplinghat and M.S. Turner have attempted 
to derive MoND from cold dark matter, but this work
has completely ignored
the low-surface-brightness galaxies in which the acceleration is
everywhere smaller than the critical acceleration introduced in MoND.}
\footnote{Proper references to work due to other authors 
can be found in Ref. 1.}

It is well-known that cold dark matter (DM) works wonders for 
cosmology. 
However, at the galactic scale, dark matter does not fare 
nearly
as well, for (1) it can explain the observed asymptotic 
independence of orbital velocities
on the size of the orbit only by fitting data (usually with two
parameters) for individual galaxies, and (2) it cannot explain,
in a natural manner,
the observed baryonic Tully-Fisher (TF) relation, i.e., the
asymptotic-velocity-mass ($v^4 \propto M$) relation.

This is in great contrast to 
modified Newtonian dyanmics (MoND), a scheme
due to M. Milgrom.  MoND stipulates that the
acceleration of a test mass $m$ due to the source $M$ is given by
$a = a_N$ for $a \gg a_c$, and $a = \sqrt{a_N\,a_c}$ for 
$a \ll a_c$,
where $a_N= G M /r^2$ is the usual Newtonian acceleration and the 
critical acceleration
is given by $a_c \sim 10^{-8} cm/s^2$
($\approx c H/(2 \pi)$ numerically, 
$H$ being the Hubble parameter).
With only one parameter (viz., $a_c$)
MoND can explain fairly successfully the observed
flat galactic rotation curves and the observed TF relation.
But admittedly there are problems with MoND at the cluster and 
cosmological scales.

Thus DM and MoND complement each other well, each being successful
where the other is less so. It is natural for us
to combine their salient successful features
into a unified scheme 
--- all within the framework of quantum gravity.\cite{HMN}
Significantly we find that (1) 
MoND is a phenomenological consequence of quantum gravity, with the 
critical galactic acceleration $a_c$ correctly predicted to have
magnitude
$\sqrt{\Lambda / 3} \sim H$ ($\Lambda$ denotes the cosmological 
constant);
(2) the MoNDian force law, at the galactic
scale, is merely a manifestation of DM; (3)
global physics ($\Lambda$) and local galactic dynamics 
($a_c$) are connected; (4) the mass profile of DM is related
to the energy contents due to
ordinary matter and dark energy ($\Lambda$); 
(5) in the formulation of MoNDian dark matter (MDM) via
gravitational Born-Infeld theory, the quanta of MDM 
are shown to obey infinite statistics (IS), like the 
``particles'' constituting dark energy as proposed in the
holographic-quantum-foam-inspired 
cosmology\cite{ng}\footnote{Thus, the quanta
constituting the dark sector obey infinite statistics, rather than
Fermi or Bose statistics.  This may be the main
difference between dark energy/matter and ordinary matter.
}.

\section{Reinterpretation of MoND via Eric Verlinde's  
Entropic Gravity}

Noting that Newton's law
of gravity is precisely the fundamental relation that Milgrom
proposes to modify so as to fit galactic rotation curves \&  TF
relation, we choose to start with
Verlinde's recent proposal of entropic 
gravity (based in part on Jacobson's idea that gravity is just an 
effect of bodies increasing their entropy.)  

First let us review
how Verlinde rederives the canonical Newton's laws.

(I) Succintly 
Verlinde derives Newton's 2nd law
$
\vec{F} = m \vec{a},
$\,
by using
(a) the first law of thermodynamics to propose the concept of
entropic force
$
F_{entropic} = T \frac{\Delta S}{\Delta x},
$
and invoking Bekenstein's original arguments
concerning the entropy $S$ of black holes:
$
\Delta S = 2\pi k_B \frac{mc}{\hbar} \Delta x
$; and
(b) the formula for the Unruh temperature,
$
k_B T = \frac{\hbar a}{ 2\pi c},
$\,
associated with a uniformly accelerating (Rindler) observer.

(II). Next Verlinde rederives Newton's law of gravity $a= G M /r^2$\, 
by considering an
imaginary quasi-local (spherical) holographic screen of area $A=4 \pi 
r^2$ with temperature $T$, and using
(a) Equipartition of energy $E= \frac{1}{2} N k_B T$
with $N = Ac^3/(G \hbar)$ being
the total number of degrees of freedom (bits) on the screen; and
(b) the Unruh
temperature formula and the fact that $E= M c^2$.

Since we live in an accelerating universe, let us
generalize Verlinde's proposal to de
Sitter (dS) space with positive
cosmological constant $\Lambda$.  In such a dS space,  
the Unruh-Hawking temperature, as measured by an inertial observer,
is $T_{dS} = \frac{1}{2\pi k_B} a_0$ where 
$a_0=\sqrt{\Lambda / 3}
\sim H$ (numerically). 
The \emph{net} temperature as measured by the non-inertial observer 
(due to some
matter sources that cause the acceleration $a$\,) is \,$\tilde{T}\equiv
T_{dS+a}-
T_{dS}
=\frac{1}{2\pi k_B} [\sqrt{a^2+a_0^2} - a_0]$.
Part (I) of Verlinde's approach can now be generalized to yield the 
entropic force (in deSitter space) 
$F_{entropic}= \tilde{T}\, \nabla_x S= m 
[\sqrt{a^2+a_0^2}-a_0]$.
For $ a \gg a_0$, we have $F_{entropic}\approx ma$. For
$a \ll a_0$:
$
F_{entropic}\approx m \frac{a^2}{2\,a_0},
$
so
the terminal velocity $v$ of the test mass $m$
should be determined from
\,$ m a^2/(2a_0) = m v^2/r$.
In the small acceleration $a \ll a_0$ regime,
the observed flat galactic rotation curves ($v$ is independent of $r$) 
and the observed TF relation 
now require (recall $a_N = GM/r^2$) that
$ a \approx \left(\,2 \, a_N \,a_0^3 \,/\pi \right)^{\frac14}$.
But that means
$F_{entropic} \approx m \frac{a^2}{2\,a_0} = F_{Milgrom} \approx m
\sqrt{a_N a_c}\,$.
Thus we have recovered MoND --- provided we identify 
$a_0 \approx 2 \pi a_c$, with the critical
galactic acceleration $a_c \sim \sqrt{\Lambda/3} \sim H \sim 10^{-8} 
cm/s^2$.
From our perspective, MoND is a phenomenological
consequence of quantum gravity; furthermore we have correctly predicted 
the magnitude of $a_c$!

Part (II) of Verlinde's argument is straightforwardly generalized to
give
$2 \pi k_B \tilde{T} 
= \frac{G\,\tilde{M}}{r^2}$,
where $\tilde{M} = M + M'$ represents the \emph{total} mass enclosed 
within the volume $V = 4 \pi r^3 / 3$, with
$M'$ being some unknown mass, i.e., dark
matter.  Now consistency demands that 
$M'= \frac{1}{\pi}\,\left(\,\frac{a_0}{a}\,\right)^2\, M$;
that is,
$F_{entropic} = m [\sqrt{a^2+a_0^2}-a_0] = m\,a_N
\left[\,1+ (a_0/a)^2/\pi \right]$.
For $a \gg a_0$, we recover the Newtonian force law $ F_{entropic} 
\approx m a \approx m a_N$, and hence
$a=a_N$ (and $M' \approx 0$).  But
for $a \ll a_0$, we have
$ F_{entropic} \approx  m \frac{a^2}{2\,a_0} \approx m a_N (1/\pi)
(a_0/a)^2$,
yielding
$a= \left(\,2\, a_N \,a_0^3 / \pi \, \right)^{\frac14}$ (as required)
and the dark matter mass profile $M' \sim 
(\sqrt{\Lambda}/G)^{1/2}M^{1/2}r$.

The DM-MoND duality is now apparent:
On one hand, we can interpret the $F_{entropic}$ equation 
to mean that there is
\emph{no} dark matter,
but that the law of gravity is modified (according to MoND).
On the other hand, we can rewrite $F_{entropic}$ equation as
$F_{entropic}= mG(M+M')/r^2$ ,
with $M'$ denoting the dark matter
which, by construction, is compatible with MoND. At
galactic scales, $M', \Lambda$ and $M$ are related to one another.
Dark matter of this kind can behave \emph{as if} there is 
no dark matter but MoND.  Therefore, we call it ``MoNDian dark 
matter".

While the MoNDian dark matter profile given by
$M' = \frac{1}{\pi}\,\left(\,\frac{a_0}{a}\,\right)^2\, M\,$
reproduces the correct force laws (to the leading order)
in both regimes of $a \gg a_0$ and $a \ll a_0$, we expect      a 
more generic profile of the form
$M' = \left[\,\lambda a_0/a +
(a_0/a)^2/\pi \right] \, M\,$,
with $\lambda >0$ (and of order 1) which ensures that $M' > 0$ when $a 
\gg a_0$.
As a function of $r$, the dark matter profile now reads (for 
$a\gg a_0$):
$M' \approx \left[\, \lambda \,\left(a_0 /a_N \right) + 
\left( 
\,\frac{1}{\pi} - \lambda
\,(1 + \lambda) \,\right)\, \left( a_0/a_N \right)^2 
\,\right] \,M \,$.
In principle, this mass profile can be checked against 
observational data. 

We conclude this section with an observation on the application of
MoNDian dark matter to cosmology.\cite{HMN}  One can in principle
have Einstein's gravity together with a MoNDian dark matter source.
The departure from MoND happens when we replace $\tilde{M}$
with the
active gravitational (Tolman-Komar) mass,
i.e. when a non-relativistic source is
replaced by a fully relativistic source. In that case, we have
$\sqrt{a^2+a_0^2}-a_0  = \frac{G\, (\,M(t)+M'(t)\,)}{\tilde{r}^2} + 4
\pi G
\,p\,
\tilde{r} -\frac{\Lambda}{3}\,\tilde{r}$,
where $\tilde{r}$
is the physical radius.  If we naively
use MoND at the cluster scale, we would be
missing the pressure ($p$) and cosmological constant terms 
$4 \pi G \,p\, \tilde{r} -\frac{\Lambda}{3}\,\tilde{r}$\,
which could be significant. This may explain why MoND
doesn't work well at the cluster scale,
despite the DM-MoND duality realized at the galactic scale.

\section{MoNDian Dark Matter via Gravitational Born-Infeld Theory}

A particularly useful reformulation of MoND is via an effective
gravitational dielectric medium, motivated by the analogy 
between
Coulomb's law in a dielectric medium and Milgrom's law for MoND.
\footnote{It has been known to L. Blanchet, Milgrom and others that
the MoNDian force law can be formulated as being governed by a 
nonlinear generalization of Poisson's equation which describes
the nonlinear electrostatics embodied in the Born-Infeld theory.}
Starting from the Born-Infeld theory of electrostatics, we
can write  
the corresponding gravitational Hamiltonian density as
$H_g = b^2 \left(\,\sqrt{1+ 
D_g^2/b^2}-1\,\right)/(4 \pi)$,
where $D$ stands for the displacement vector and $b$ is the maximum
field strength in the Born-Infeld theory.
With $A_0 \equiv b^2$ and $ \vec{A} \equiv b \, \vec{D_g}$, the 
Hamiltonian density becomes
$H_g = \left(\,\sqrt{A^2+A_0^2}-A_0\,\right)/(4 \pi)$. 

As in the Verlinde approach, let us
assume energy equipartition. Then the 
effective gravitational Hamiltonian density is equal to
$H_g = \frac{1}{2}\,k_B\, T_{\rm eff}\,$.
The Unruh temperature formula $T_{eff} = 
\frac{\hbar}{2\,\pi\,k_B}\, a_{\rm eff}\,$
implies that the effective acceleration is given by
$a_{eff}  = \sqrt{A^2+A_0^2}-A_0\,$.
Next we make use of the 
equivalence principle which suggests that we should identify
(at least locally) the local accelerations $\vec{a}$ and $\vec{a}_0$
with the local gravitational fields
$\vec{A}$ and $\vec{A}_0$ respectively, viz.,
$\vec{a} \equiv \vec{A}, \quad \vec{a}_0 \equiv \vec{A}_0\,$. Then 
$a_{\rm{eff}}$
should be identified as
$a_{\rm eff} \equiv
\sqrt{a^2+a_0^2}-a_0\,$,
which, in turn, implies that the Born-Infeld inspired force law takes 
the form (for a given test mass $m$)
$F_{\rm BI} = m\, \left(\,\sqrt{a^2+a_0^2}-a_0\,\right)\,$,
which is precisely the MoNDian force law!

To be a viable cold dark matter candidate, the quanta of
our MoNDian dark matter are expected
to be much heavier than $k_B\,T_{\rm{eff}}$.  Now
recall that the equipartition theorem in general states that
the average of the Hamiltonian is given by
$\langle H \rangle = - \frac{\partial \log{Z(\beta)}}{\partial 
\beta}\,$,
where $\beta^{-1} = k_B T$ and $Z$ denotes the partition function. To 
obtain
$\langle H \rangle = \frac{1}{2} \,k_B\, T$ per degree of freedom, 
even for very low temperature,
we require $Z$ to be of the Boltzmann form
$Z = \exp(\,-\beta\, H\,)\,$.
\footnote{Note that the conventional quantum-mechanical
Bose-Einstein or Fermi-Dirac statistics
would not lead to $\langle H \rangle = \frac{1}{2}\, k_B\, T$
per degree of freedom at low temperataure.}
But this is precisely what is called the infinite
statistics \footnote{Notable properties of infinite statistics
include:
particles obeying IS are virtually
distinguishable;
the partition function is
$Z = \Sigma e^{- \beta H}$, with NO Gibbs factor;
in IS, all representations of the particle permutation
group can occur;
theories of particles obeying IS are non-local;
TCP theorem and cluster decomposition still hold; and 
quantum field theories with IS remain unitary.}
as described by the Cuntz algebra
(a curious average of the bosonic and fermionic algebras)
$a_i \, a^{\dagger}_j = \delta_{ij}\,$.
It is intriguing that 
only by invoking infinite statistics for the
microscopic quanta which underly the thermodynamic description
of gravity implying such a MoNDian force law,
can the assumption of energy 
equipartition,
even for very low temperature $T_{\rm{eff}}$, be justified.




\bibliographystyle{ws-procs975x65}
\bibliography{ws-pro-sample}

\end{document}